\documentclass[11pt,a4paper]{article}

\usepackage{epsfig}
\usepackage{amsmath}
\usepackage{amssymb}
\usepackage{verbatim}
\usepackage{bm}
\usepackage{dsfont}
\usepackage[footnotesize]{caption}
\usepackage{subfigure}
\usepackage{cite}
\usepackage{textcomp}
\usepackage{calc}
\usepackage{geometry}
\usepackage{bbm}
\usepackage{xcolor}
\usepackage{multirow}
\usepackage{color}
\usepackage{float}
\usepackage{bbold}
\usepackage{ulem}

\usepackage{hyperref}
\hypersetup{colorlinks=true,urlcolor=blue,linkcolor=red,citecolor=green!60!black}

\renewcommand{\baselinestretch}{1.3}

\begin{document}


\thispagestyle{empty}

\noindent 
\hfill 

\vskip 1.5cm

\begin{center}
{\LARGE\bf Is the radion a bulk Higgs doublet?}

\vskip 1cm

\renewcommand*{\thefootnote}{\fnsymbol{footnote}}

{\large
Marco Merchand,$^{a,\,\hspace{-0.25mm}}$%
\footnote{mamerchandmedi@email.wm.edu}
Marc Sher$^{a,\,\hspace{-0.25mm}}$%
\footnote{mtsher@wm.edu}
and Keith Thrasher,$^{a,\,\hspace{-0.25mm}}$%
\footnote{rkthrasher@email.wm.edu}
}\\[3mm]
{\it{
$^{a}$ High Energy Theory Group, College of William and Mary, Williamsburg, \\
 VA 23187, USA}}

\end{center}

\vskip 1cm

\renewcommand*{\thefootnote}{\arabic{footnote}}
\setcounter{footnote}{0}


\begin{abstract}


In 2013, Geller, Bar-Shalom and Soni (GBS) proposed a Randall-Sundrum model with a bulk SU(2) doublet.  They found that the doublet could both stabilize the extra dimension and break the electroweak symmetry; the radion and the Higgs boson were then one and the same.  This alternative to the Goldberger-Wise mechanism contained enhanced Higgs couplings to gluons and photons, but at the time of the GBS paper, these couplings were phenomenologically acceptable. We note that updated results from the LHC rule out the model as presented.   One can expand the parameter space by extending the model to include bulk fermions.   We do so, and show that for any choice of parameters, the model remains in conflict with LHC data.    We then extend the scalar sector to a two Higgs model, and show that the model contains a phenomenologically unacceptable zero mass scalar.  This property will likely apply  to all extensions to the scalar sector, and thus we conclude that the radion cannot be part of a bulk Higgs doublet.


\end{abstract}


\newpage

{\hypersetup{linkcolor=black}\renewcommand{\baselinestretch}{1} \tableofcontents}


\section{Introduction}
\label{sec:introduction}


 Theories with extra spacetime dimensions provide an elegant solution to the hierarchy problem. In the model proposed by Randall and Sundrum (RS) \cite{Randall:1999ee} the universe has five spacetime dimensions with the extra dimension being spatial and compactified into an orbifold $S_1/Z_2$ with two $4D$ branes fixed at $y=0$, the UV brane, and at $y=y_c$, the IR brane, where $y$ is the coordinate of the extra dimension. The SM is placed in the IR brane and an exponential hierarchy is then generated by gravitational red shifting, i.e., $M_{\text{TeV}} = M_{Pl} e^{- k y_c} $ therefore requiring $k y_c \approx 37$ where $k$ is the curvature scale and is of order the Planck scale. 

  The RS model has the theoretical shortcoming that the value of the brane separation, $y_c$, is not fixed by any dynamical mechanism and in the 4D effective field theory gives rise \cite{Goldberger:1999un} to a massless scalar field. This scalar field, associated with fluctuations in the size of the extra dimension is called the radion or, by its ADS/CFT dual \cite{Rattazzi:2000hs}, the dilaton. A stabilization mechanism that gives mass to the radion was devised by Goldberger and Wise (GW) \cite{Goldberger:1999uk} by introducing a bulk scalar field with appropriate bulk and brane potentials. Taking the quartic interactions of the scalar on the brane to be large, the dynamics of this scalar generate an effective potential for the radion and the global minimum of the potential fixes the size of the extra dimension without fine tuning of the parameters. In the GW mechanism the authors did not take into account the backreaction of the bulk scalar on the background geometry. A technique for generating solutions to the coupled scalar-gravity Einstein equations was presented in \cite{DeWolfe:1999cp} and a full account of these effects together with a perturbative calculation for the radion mass and couplings is given in the paper of Csaki et al. \cite{Csaki:2000zn}. 
  
  In the original RS model only gravity propagates in the bulk of the extra dimension  \cite{Randall:1999ee, Davoudiasl:1999jd}. It was later shown \cite{Grossman:1999ra, Gherghetta:2000qt, Pomarol:1999ad} that allowing the SM fermions and gauge bosons to propagate in the bulk could offer a natural explanation of the flavor hierarchy. 
  
  The scenario of a bulk Higgs was discussed in \cite{Chang:1999nh, Huber:2000fh, Davoudiasl:2000wi}. In this references the authors assumed that the $5D$ bulk Higgs was responsible for inducing EWSB by a constant VEV. This scenario required an extreme fine-tuning of parameters in order to generate $\mathcal{O}(\text{TeV})$ gauge boson masses, a problem that the RS model was purported to resolve. Thus it was concluded that the Higgs had to remain on the TeV brane in order to avoid the hierarchy problem.

Later on, in Ref. \cite{Davoudiasl:2005uu}, the authors did a careful treatment of EWSB by a $5D$ bulk Higgs and derived the conditions necessary for the appearance of a tachyonic zero-mode such that a realistic $4D$ picture of EWSB can be achieved. The Higgs KK zero-mode develops a non-trivial profile function and is no longer a constant. This is different from the two references above which assumed a constant VEV. The authors showed that no fine-tuning is necessary and provided a phenomenologically acceptable scenario of a universal extra dimension (UED). Models with a bulk Higgs have been studied extensively after that, see Ref. \cite{Cacciapaglia:2006mz, Archer:2012qa, Archer:2014jca, Haba:2009wa, Quiros:2013yaa, Mahmoudi:2016aib}.

Several studies motivated by the question of whether the SM Higgs can  be the GW radion of the RS model  have appeared in the literature.
In  \cite{Haba:2011tc} the authors studied the profile function $\phi(y)$ of a bulk scalar field under general boundary conditions (BCs) and showed that a bulk scalar with non-zero Dirichlet BCs $(D,D)$ on the branes and without brane potentials is equivalent to the GW mechanism and can therefore stabilize the radius. They concluded that a bulk SM Higgs cannot be the GW stabilizer unless one assumes unnaturally small values of its bulk mass, but that a Higgs triplet under the $SU(2)_R$ gauge group of the custodial model in \cite{Agashe:2003zs} can provide stabilization.

Since the radion field emerges as the lightest new state, there is the possibility of it being experimentally accessible.   Particular attention has been placed on the curvature-Higgs term $\xi\mathcal{R}\  \Phi^\dagger \Phi$ since after expanding about the radion field VEV this term induces kinetic mixing between the radion field and the Higgs, therefore requiring a non-unitary transformation to obtain the canonically normalized degrees of freedom. After diagonalization the physical fields become mixtures of the original non-mixed radion and Higgs boson. The phenomenological consequences of a non-zero mixing  have been studied extensively in the literature  \cite{Csaki:2000zn, Giudice:2000av, Hewett:2002nk, Dominici:2002jv,  Chaichian:2001rq, Datta:2003kf, deSandes:2011zs, Kubota:2012in, Desai:2013pga, Boos:2014xha, Boos:2015xma, Frank:2016oqi, Ahmed:2015uqt}
 
  It was later shown by Geller, Bar-Shalom and Soni (GBS) \cite{Geller:2013cfa} that the GW mechanism, as it is, cannot be applied to an $SU(2)_L$ doublet stabilizer since $(D,D)$ BCs yield a phenomenologically unacceptable value of the EW scale and therefore one has to adopt a different choice for applying the BCs. They showed that the condition $\phi(0) \ll M_{\text{Pl}}$ necessary to generate the correct value of the EW scale does not add to the level of fine tuning already required in order to have a vanishing cosmological constant. They solved the coupled scalar-gravity system of equations by assuming a small backreaction of the $SU(2)_L$ doublet stabilizer on the metric. In \cite{Ahmed:2015mva} a complete account of the backreaction was calculated using the superpotential method of Ref. \cite{DeWolfe:1999cp}. A similar calculation was presented in  \cite{Egorov:2017vcs}.
  
  In Ref. \cite{Vecchi:2010em} it was argued that the stabilization of the extra dimension by a bulk Higgs is possible, provided the $5D$ bulk mass satisfies the Breitenlohner-Freedman bound, which corresponds to $m^2 = -4 k^2$. And as we will see this is consistent with the GBS condition $\nu \sim 1$. In this reference the author provides a qualitative analysis of the stabilization mechanism from the CFT side of the correspondence.
  
  The work of GBS provided an alternative to the GW mechanism.    They considered a bulk doublet, with different boundary conditions than GW, and showed that this single doublet could both stabilize the extra dimension and break the electroweak symmetry.    Thus, in this model the Higgs and radion are one and the same (which they refer to as a ``Higgs-radion").      Since the couplings of the radion to gluons and photons are enhanced, one has the couplings of the Higgs to gluons and photons enhanced.    At that time of their work, not only were these enhancements phenomenologically acceptable, but there did appear to be an enhancement in the diphoton decay of the Higgs.   In this paper, we first summarize the GBS model and show that current LHC data rule the model out.   
  
  In the GBS paper the fermions were allowed to propagate in the bulk, (with the exception of the top quark whose right-handed profile was given by a delta function peaked at the IR brane) however their profiles were assumed to be constants along the extra dimension. Relaxing this assumption, we investigate the values of the $5D$ bulk fermion masses that reproduce the correct hierarchy, thus adding an extra free parameter, and still show that LHC data excludes the model.    Finally, we consider a substantial expansion of the parameter-space by looking at a Two Higgs Doublet Model, and find that the model will contain a physical, zero-mass scalar, which is phenomenologically excluded.   The presence of such zero modes in multiple bulk scalar models had previously been demonstrated for soft-wall models by George \cite{George:2011tn}, but it appears that the result applies to GBS-type models as well.    Thus, it appears that the attractive idea of having the same field responsible for radius stabilization and electroweak symmetry breaking appears to be excluded.    In Section II, we describe the GBS model and show that current LHC data exclude the model as originally presented, in Section III we extend the model to include bulk fermions, and in Section IV we describe the two-Higgs generalization of the model.   Section V contains our conclusions.


\section{Higgs-Radion Unification}
\label{GBS model}
In the model of Geller, Bar-Shalom and Soni  \cite{Geller:2013cfa}, the possibility of a bulk scalar $SU(2)_L$ doublet which can stabilize the radius of the warped RS model and provide the source of electroweak symmetry breaking leading to a unified Higgs-radion state was presented. This Higgs-radion serves as an intriguing alternative to the usual radius stabilization via the GW mechanism. However, as we will see,  some of the phenomenological signatures predicted by this model are now at odds with recent LHC data, particularly the combined ATLAS and CMS measurement   \cite{ATLAS:2018doi, Khachatryan:2016vau} of $BR(H_{\text{SM}} \rightarrow \gamma \gamma)/BR(H_{\text{SM}} \rightarrow ZZ)$.

The model is the standard RS framework with two branes located at the orbifold fixed points $y=0$ and $y=y_c$. The $5D$ metric is  \begin{equation}
 ds^2 = e^{-2 A} \eta_{\mu \nu} dx^\mu dx^\nu -dy^2 , \label{metric}
 \end{equation}
 where $A$ is a function of the extra dimension that is determined by solving Einstein equations. As in the GW mechanism, the bulk scalar has both a bulk potential and potentials sourced on each of the two branes. The actions of the bulk-brane system are,
 \begin{equation}
 S_{\text{Bulk}}  = \frac{1}{2} \int d^4x \int dy \sqrt{g}  \left( g^{MN} \partial_M \Phi \partial_N \Phi   - V(\Phi)  +  6\frac{k^2}{\kappa^2}  \right),
 \end{equation}
 \begin{equation}
 S_{\text{Brane}} = - \int d^4 x  \sqrt{g_4} \ V_i^{\text{Brane}}(\Phi),
 \end{equation}
 where 
  \begin{equation}
V(\Phi) = m^2 \Phi^2,
\end{equation} 
 \begin{equation}
 V_i^{\text{Brane}}(\Phi) = \lambda_i \Phi^4 + m_i^2 \Phi^2  + \Lambda_i,
 \end{equation}
\begin{equation}
\kappa^2 = \frac{1}{2M^3}.
\end{equation}
 Here the subscript $i=$Planck, TeV is used to denote each of the two branes, $g^{MN}$ represents the $5D$ metric tensor and $g_i^{\mu \nu}$ are the induced metrics on each brane.
 From the Einstein and Euler-Lagrange equations one obtains the following
  \begin{equation}
 4 A'^2- A'' = 4 k^2 - \frac{2 \kappa^2}{3}V(\phi_0) - \frac{ \kappa^2}{3} V^{\text{Brane}}_i(\phi_0)\delta(y-y_i),   
 \end{equation}  
   \begin{equation}
 A'^2 =  k^2 + \frac{\kappa^2 \phi_0'^2}{12} - \frac{\kappa^2}{6} V(\phi_0),  \label{Geinstein2}
  \end{equation}
 \begin{equation}
 \phi_0'' = 4 A' \phi_0'  + \frac{\partial V( \phi_0 )}{\partial \phi_0} + \frac{\partial V_i^{\text{Brane}}(\phi_0)}{\partial \phi_0}\delta(y-y_i),
 \end{equation}
where primes denote derivatives with respect to $y$ and it is assumed that VEV profile of the bulk field depends only on the extra dimension. By matching the singular terms in the previous equations, the following boundary conditions may be obtained:
\begin{equation}
[ A'  ]_i = \frac{\kappa^2}{3} V_i^{\text{Brane}}(\phi_0), \quad   [ \phi_0' ]_i = \frac{\partial V_i^{\text{Brane}}(\phi_0)}{\partial \phi_0}. \label{GBC}
\end{equation}
 There are three relevant integration constants in this set of equations, two for $\phi_0$ and one for $A$. However the constant in $A$ is completely irrelevant by scale invariance of the metric. The last constant is the separation between the branes  $y_c$. With the four boundary conditions above one can determine three of the integration constant leaving one of the conditions to be tuned. This fine-tuning is the inescapable one due to the cosmological constant.
 In the GBS model the authors considered the small backreaction ($l^2 \ll 1$) approximation, such that $A=  k y + \mathcal{O}(l^2)$ and the Euler-Lagrange equations for the background VEV simplifify to 
\begin{equation}
\phi_0''= 4 k \phi_0' + \frac{\partial V(\phi_0)}{\phi_0}, 
\end{equation}
 which has the most general solution
 \begin{equation}
 \phi_0 = e^{2 k (y-y_c)} \left( C_1 e^{\nu k (y-yc)} + C_2 e^{-\nu  k (y-y_c)}  \right),
 \end{equation}
  where $\nu = \sqrt{4  + m^2/ k^2} $, similar to the GW mechanism and $C_1$ and $C_2$ are integration constants. If one is going to identify the bulk doublet with a $5D$ Higgs field that is responsible for EWSB and to give the electroweak gauge bosons their masses one has to impose that the effective $4D$ VEV, given by 
 \begin{equation}
 v_{eff}^2 =  \int dy \ \phi_0^2 \ e^{-2A},
 \end{equation}
 is of order the EW scale. This implies that one must tune the value of the Higgs profile at the Planck brane to be $\phi_0(0) \ll M_{\text{Pl}}^{3/2}$. One can achieve this by requiring $\nu <1$ so that $\phi_0 \approx C_2 e^{(2-\nu)k(y-y_c)}$ is near the Planck brane  leading to a small value  as desired. This is one of the key differences from the GW scenario where $\nu \sim 2$ giving order Planck values for the VEV on both branes which, as was argued, cannot work for a $SU(2)_L$ stabilizer. The condition $\nu <1$ implies $m^2 \approx -4 k^2$ which corresponds to the Breitenlohner-Freedman bound, consistent with Ref. \cite{Vecchi:2010em}. 
  The condition $\phi_0(0) \ll M_{\text{Pl}}^{3/2}$ forces one to choose different boundary conditions from those used in the GW stabilization. In this case one chooses the two conditions in \eqref{GBC} for the values of $\phi_0'$ at the two branes, together with the condition of the value of $A'$ at the TeV brane to solve for  $C_1$, $C_2$ and $y_c$. No additional fine tunings, beyond that needed to fine-tune the cosmological constant, have been introduced to stabilize the extra dimension.
 
 Once the solutions for the background equations are found one has to consider the scalar perturbations of the metric and the bulk doublet and solve the linearized Einstein equations to find the spectrum of physical fields. These perturbations are parametrized as \cite{Charmousis:1999rg}
\begin{equation}
\Phi_i(x,y) = \begin{pmatrix}
0 \\
\phi_0(y) + \varphi(x,y)
\end{pmatrix}, \label{bulkdoublet}
\end{equation}
\begin{equation}
ds^2 = e^{-2A-2F(x,y)} \eta_{\mu \nu} dx^\mu dx^\nu - (1+2F(x,y))^2 dy^2.  \label{background}
\end{equation} 
 and the linearized Einstein equations are given \cite{Csaki:2000zn}
\begin{equation}
F'' - 2 A' F' - 4 A'' F - 2 \frac{\phi_0''}{\phi_0'}F' + 4 A' \frac{\phi_0''}{\phi_0'}F = e^{2A} \Box F,
\end{equation}
 \begin{equation}
\phi_0'\varphi= \frac{3}{\kappa^2}\left(F'-2A'F \right),
 \end{equation}
 where the second equations imply that the scalar perturbations can be KK decomposed as 
 \begin{eqnarray}
\varphi(x,y)&=&\sum \varphi_n(y)h_n(x) \label{KKphi} ~,\\
F(x,y)&=&\sum F_n(y)h_n(x) \label{KKF} ~.
\end{eqnarray}
The boundary conditions are given by 
 \begin{equation}
\left[\varphi' \right]_i=\frac{\partial^2 V^{Brane}_i(\phi_0)}{\partial \phi^2} \varphi + 2 \frac{\partial V^{Brane}_i(\phi_0) }{\partial \phi}F
\label{BCF1}~,
\end{equation}
 and the stiff boundary potential assumption cannot be made here. This is different from the GW case where one takes the quartic coupling to be infinitely large in order to simplify the boundary conditions. The resulting mass and effective VEV of the so called "Higgs-radion" field were found to be given to the lowest order in the backreaction by 
 \begin{equation}
 m_{h_r}^2 = l^2 \frac{52 k^2}{15 k y_c} e^{2 k y_c}, 
 \end{equation}
 \begin{equation}
 v_{eff}^2 = l^2 \frac{2}{5 k \kappa^2}e^{2 k y_c}.
 \end{equation}
 The solutions for the scalar fluctuations can be written in terms of the canonically normalized Higgs-radion field and to the lowest order are 
 \begin{eqnarray}
F(x,y) &=& h_r\frac{e^{2k(y-y_c)}}{\Lambda_r}  ~, \\
\varphi(x,y) & \approx & h_r\frac{1}{\Lambda_r e^{2ky_c}} \left( 2\sqrt{5} e^{4ky} \sqrt{\frac{1}{\kappa^2}}-\frac{13 e^{2 k y + 2 k y_c} \sqrt{\frac{1}{\kappa^2}} y}{\sqrt{5} y_c} \right) \ell  ~,  \label{profgeller}
\end{eqnarray}
\begin{equation}
l^2 \equiv \frac{1}{4} e^{-2(2-\nu)k y_c} \kappa^2 \phi_{TeV}^2 ( 20 - 4 \nu -  3 \nu^2)  
\end{equation}
 where $l^2$ parametrizes the backreaction and $\Lambda_r \equiv \sqrt{6}M_{\text{Pl}}e^{-k y_c}$ is the canonical normalization. Since the Higgs-radion field arises from the massive zero mode of both the metric and scalar perturbations, its couplings to the SM matter fields are different from those of the GW pure radion state. Thus one has to add the contributions coming from the $5D$ Higgs couplings to those coming from the conformal couplings of the radion. In paticular, due to the trace anomaly, the radion has couplings to photons proportional to the QED beta function in addition to the usual momentum dependent effective couplings coming from fermions and the W gauge boson running around loops. The Higgs scalar only has the latter contributions.  A table with the most relevant couplings is given in table \ref{hr couplings}. These couplings were calculated by GBS, we refer the reader to section IV of Ref. \cite{Geller:2013cfa}. The somewhat surprising perfect integers of $3$ and $9$ in the couplings to fermions and gauge bosons arise after expanding the Higgs kinetic term and  integrating the product of profiles for the vev function and scalar perturbation. To lowest order in the backreaction, the gauge bosons have a constant profile $V_\mu(x,y) = 1/\sqrt{y_c}V_\mu^{(0)}(x)$ and its mass scales quadratically with the backreaction $l$, allowing us to express the coupling in terms of the gauge boson masses, as it is usually the case. Corrections to these formulas are order $\mathcal{O}(l^4)$ and will be neglected here.

In the GBS model the top quark profile was assumed to be a delta function peaked at the TeV brane while the lighter fermions profiles were approximated to be flat. Setting the Higgs-radion mass to $m_{h_r} = 126 $ GeV and using the three level couplings of table \ref{hr couplings}, the branching fractions to different decay channels may be calculated, see their Table II in  \cite{Geller:2013cfa} where they compared with the SM Higgs values. 

 The most significant differences between the Higgs-radion and the SM Higgs come from the branching fractions to photons and gluons. Recent ATLAS measurements of the Higgs production and decay modes at $80 \ \text{fb}^{-1}$ of luminosity \cite{ATLAS:2018doi}, obtained $BR(H \rightarrow \gamma \gamma)/BR(H \rightarrow ZZ) = 0.076 ^{+0.013}_{-0.011} $ for which the corresponding values in Table II in  \cite{Geller:2013cfa} yield the ratio $BR(h_r \rightarrow \gamma \gamma)/BR(h_r \rightarrow ZZ) = 0.147$.  However, as GBS mentioned, this is a leading order calculation that does not take into account the effects of summing over the one-loop contributions 
of the KK towers. In Ref. \cite{Archer:2014jca}, the effect of including these corrections from the KK towers of fermions were investigated and it was shown that for a low KK scale, $M_{KK} \approx 3.3 \ $ TeV there is a sizable suppression in the $h \rightarrow \gamma \gamma $ decay width relative to the SM Higgs decay width, see their figure $8$.

From fitting the Higgs-radion normalization $\Lambda_r$ to the Higgs signal strengths data, the GBS authors found that the KK scale must lie in the range  $4.48 $ TeV $ < M_{KK} < 5.44 $ TeV. In Ref \cite{Archer:2014jca} the authors calculated the decay rate $H \rightarrow \gamma \gamma$ for a bulk Higgs including the corrections coming from the KK towers of fermions, charged scalars and gauge bosons. They found that the most significant deviation from the SM came from KK quarks and leptons effects unless one assumes unnaturally small values for the Yukawa couplings. 
As can be seen from their figure 8 (bottom-left) the deviations from the SM are less than $5\%$ for Yukawa couplings $y=1,2$ in the range of KK scales that are allowed in the GBS model. Also in the same reference, it was found that the one-loop corrections from KK towers on $h \rightarrow W W^*$ would correspond to a less than $1\%$ supression. 

In the GBS model the Higgs-radion is expected to have significant deviation in its couplings to fermions relative to the bulk Higgs of Ref. \cite{Archer:2014jca} so that KK towers effects might be more pronounced. Even then it would be challenging to accommodate the Higgs branching fraction within experimental levels.  
 \begin{center}
 \begin{table}
\makebox[1 \textwidth][c]{ \scalebox{1.1}{
\begin{tabular}{| c | c |}
\hline
 & $h_r$ Couplings     \\  \cline{1-2}
 $h_r tt$&  $4 \frac{m_t}{\Lambda_r}$ \\ \cline{1-2}
 $h_r \bar{f} f$ for $f \neq t$ &  $\frac{9 m_f}{\Lambda_r}$    \\ \cline{1-2}
  $h_r VV$  &  -$\frac{9 m_V^2}{\Lambda_r}$    \\ \cline{1-2}
$h_r gg/\gamma \gamma$ & $ \frac{1}{\Lambda_r} \frac{\beta_{QCD/QED}}{2 g} $   \\ \cline{1-2}
\end{tabular}}}
\caption{ Couplings of $h_r$ to gauge bosons and fermions} \label{hr couplings}
\end{table}
\end{center}

\section{Bulk Fermions}
\label{Bulk Fermions}

In this section we relax the assumption made in the GBS paper that the light fermions have a constant profile along the extra dimension and we will find the parameter space of the fermion bulk mass parameters $c = m/k$ that reproduce the flavor mass hierarchies. The hope would be that the expanded parameter space might allow a viable model.

For simplicity we present the results for the quarks only. A similar analysis can be carried out for leptons.
The $5D$ Yukawa interactions are given by 
 \begin{equation}
 S_{\text{Yukawa}} =- \int d^5 x \sqrt{g} \left[ y^{(5)}_{u_{ij}}  \bar{Q}^i \tilde{\Phi} \ u^j + y^{(5)}_{d_{ij}}\bar{Q}^i \Phi \ d^j  + h.c. \right] \label{S3}
   \end{equation}
where the  $y^{(5)}_{ij}$ are the Yukawa parameters and have mass dimenion $-1/2$.  The fermion masses and interactions of the Higgs-radion $h_r$ are found by expanding the doublet as in equation \eqref{bulkdoublet}, where the VEV profile is given by \cite{Geller:2013cfa}
\begin{equation}
\phi_0(y) = \phi_{TeV} e^{(2-\nu) k  (y-y_c)}.
\end{equation}
Since fermions propagate in the bulk, the 5D Yukawa couplings can give rise to mixing between different modes and could lead to potentially dangerous FCNC's mediated by the Higgs-radion or its higher dimensional excitations. These effects can impose tight lower bounds on the KK scale \cite{Gherghetta:2000qt,Agashe:2004cp}. However we will show below that these constraints will not be necessary for our purposes and we will assume diagonal 5D Yukawa couplings.

The fermion masses are given by the integral expression
\begin{equation}
m_{u/d} = y^{(5)}_{u/d} \int_{0}^{y_c}   \phi_0(y) \chi_Q^{(0)}(y) \chi_{u/d}^{(0)}(y) dy
\end{equation}
where the zero-mode profiles where derived  in Ref. \cite{Grossman:1999ra}, to be 
\begin{equation}
\chi_Q^{(0)}(y) = \chi_Q^{(0)}(0) e^{m_Q y}, \quad  \chi_u^{(0)}(y) = \chi_u^{(0)}(0) e^{-m_u y}
\end{equation}
with normalization $\int_{0}^{y_c} dy e^{\sigma} \chi^{(m)}_\psi (y) \chi^{(n)}_\psi(y) = \delta_{mn}$ and we assumed the Yukawa constants are diagonal.
The fermion masses are then given by
\begin{equation}
m_{u,d} = y^{(5)}_{u,d}  \ \phi_{TeV}  e^{-(2-\nu) ky_c } \left\lbrace  \frac{(1+2 c^i_L)}{(e^{ k y_c (1+2c^i_L)}-1)} \frac{(1-2 c^{u,d}_R)}{(e^{ k y_c (1-2c^{u,d}_R)}-1)}\right\rbrace^{1/2} \frac{e^{A k y_c }-1}{ A } \label{S4}
\end{equation}
where  we defined $A$ and the bulk mass parameters as
\begin{equation}
A \equiv 2-\nu +c^i_L -c^{u,d}_R, \quad c^i_L \equiv \frac{m_{Q_i}}{k}, \quad c^{u,d}_R \equiv \frac{m_{u,d}}{k}.
\end{equation}

From the requirement of $v_{\text{eff}} \approx 246 $ GeV and assuming natural values for the Yukawa constants it is straightforward to find $y^{(5)} \phi_{TeV} = \sqrt{k/M_{\text{Pl}}} \ v_{\text{eff}} e^{ k y_c}$. The relation  $k/M_{\text{Pl}} \approx 1.6$ is fixed by the ratio of the Higgs-radion mass to its VEV. Also the stabilization of the extra dimension is obtained when $\frac{1}{2\nu} \sim 37$, thus we can neglect the parameter $\nu$ above. Note that for each generation, there are three mass parameters $(c_Q, c_u, c_d) $ and two masses leaving more flexibility in studying the phenomenology. The mass hierarchy of the quark sector can be explained by the profile values given in the table \ref{bulkmass}, although other choices are also viable.

Similarly the Yukawa couplings of the Higgs-radion are given by
\begin{equation}
y^{(4)}_u= y^{(5)}_u \int_{0}^{y_c} dy \chi^{(0)}_u(y) \chi^{(0)}_Q(y) \varphi_0(y),
\end{equation}
where the expression for $\varphi_0(y)$  was presented in \eqref{profgeller}. After evaluating the integral the Yukawa Lagrangian is 
\begin{equation}
-\mathcal{L}_{\text{Yukawa}} = \frac{h_r}{\Lambda_r} \left( \tilde{y}_Y m_f \bar{f}_L f_R + (KK, n>0) \right).
\end{equation}
To this couplings we have to add the gravitational contribution coming from the scalar perturbation of the metric. One finds 
\begin{equation}
-\mathcal{L}_{\text{Metric}} =  -    \frac{h_r}{\Lambda_r}  \left(  \tilde{y}_G m_f \bar{f}_L f_R + (KK, n>0) \right).
\end{equation}
and the numerical factors multiplying the masses are given in terms of the bulk mass parameters by
\begin{equation}
\tilde{y}_Y = \frac{k y_c A}{e^{ k y_c A}-1} \left\lbrace  \frac{10}{k y_c} \frac{e^{ k y_c A}-e^{-2  k y_c}}{(2+A)} - 13 \frac{e^{ k y_c A}( k y_c A-1)+1}{(k y_c A)^2}  \right\rbrace,
\end{equation}
\begin{equation}
\tilde{y}_G = \left[  2e^{-2 k y_c} \frac{e^{ (2+ A)k y_c }-1}{2 + A} \frac{A}{e^{A k y_c} - 1} \right].
\end{equation}
\begin{table}
\begin{center}
\begin{tabular}{|c|c|c|c|}
\hline
 & $c^i_L$ & $c^{u_i}_R$ & $c^{d_i}_R$  \\
 \hline
 $i=1$ & $-0.51$ &   $0.77$ & $0.74$   \\
 \hline
 $i=2$ & $0.63$ & $0.62$ & $0.7$  \\
 \hline
 $i=3$ & $0.53$ & $0.03$ & $0.58$  \\
 \hline 
\end{tabular} 
 \caption{A set of bulk mass parameters that reproduces the quark spectrum. Since each generation has three parameters and two masses, many other choices can be made. The predictions made in this section are only weakly sensitive to the specific choices; we have scanned the entire set of parameters and our conclusion in this section are unaffected} \label{bulkmass}
\end{center}
\end{table}

The overall couplings to fermions and massive gauge bosons can be written in the following Lagrangian density
\begin{equation}
\mathcal{L} = \frac{h_r}{\Lambda_r} \left( - \sum_{f} (\tilde{y}_Y- \tilde{y}_G)m_f \bar{f} f - 9 m_V^2 V_\mu V^{\mu} \right), \label{cou}
\end{equation}
where the sum is over quarks and leptons and $V_\mu V^\mu = 2 W_\mu^+ W^{\mu -}, Z_\mu Z^\mu$.

The couplings of the Higgs radion to the massless gauge bosons is zero at tree level, however, there is a loop contribution coming from the trace anomaly term
\begin{equation}
T_\mu^\mu = \sum_{a} \frac{\beta_a(g_a)}{2 g_a}F_{\mu \nu}^a F^{a \mu \nu}.
\end{equation}
The list of branching ratios for both the SM Higgs and the Higgs radion are given in table \ref{branching} where we set $\Lambda_r = 4 $ TeV
\begin{table}
\begin{center}
\begin{tabular}{|c|c|c|c|}
\hline
 $BR(h_r \rightarrow XX)$ & SM &   Higgs radion \\
 & $m_h =126 $ GeV & ($m_{h_r}=126 $GeV)  \\
 \hline
 $WW^*$ & $0.231$ &   $0.207$   \\
 \hline
 $ZZ^*$ & $0.0289$ & $0.0260$   \\
 \hline
 $gg$ & $0.0848$ & $0.183$   \\
 \hline
 $\gamma \gamma$ & $2.28 \times 10^{-3}$ & $2.47 \times 10^{-3}$   \\
 \hline
 $b \bar{b}$ & $0.561$ & $0.501$   \\
 \hline
 $\tau \bar{\tau}$ & $0.0615$ & $0.054$   \\
 \hline
 $c \bar{c}$ & $0.0283$ & $0.024$   \\
 \hline
 Total width (GeV) & $4.21 \times 10^{-3}$ & $1.36 \times 10^{-3}$   \\
 \hline
\end{tabular}
\end{center} \caption{Branching fractions to different decay channels of the SM Higgs boson and the Higgs-radion. The values of the $c$ parameters corresponds to those in table \ref{bulkmass}.} \label{branching}
\end{table}

From table \ref{branching} we can see that the predictions for the Higgs-radion branching fractions lie very close to the SM Higgs boson expectation. We scanned the parameter space of the bulk mass parameters $c=m/k$ and checked that the ratios $\frac{BR(h_r \rightarrow \gamma \gamma)}{BR (h_r \rightarrow ZZ)}$, $\frac{BR(h_r \rightarrow WW)}{BR (h_r \rightarrow ZZ)}$, $\frac{BR(h_r \rightarrow \tau \tau)}{BR (h_r \rightarrow ZZ)}$ and $\frac{BR(h_r \rightarrow bb)}{BR (h_r \rightarrow ZZ)}$ are within at least  $2 \sigma $ of the experimentally allowed region presented by the CMS and ATLAS collaborations \cite{Khachatryan:2016vau, ATLAS:2018doi}.

The fact that the radion has a narrow width allows us to correlate its production cross section with the partial width \cite{Giudice:2000av} and simply rescale by the SM Higgs cross section, i.e., $\sigma(gg\rightarrow h_r)\Gamma(H_{SM}\rightarrow gg) = \sigma(gg\rightarrow H_{SM})\Gamma(h_r \rightarrow gg)$.  The effects of both the trace anomaly and the top quark loop were included in the partial width calculation. The formula for the $h_r  \rightarrow gg$ decay rate is quoted in the GBS paper. We just inserted the appropriate quark and gauge boson couplings.
 
 The prediction for the production cross section via gluon-gluon fusion (ggF) of a Higgs-radion gives 
\begin{equation}
\sigma(gg \rightarrow h_r) \geq140 \ \text{pb}
\end{equation}
for a bulk mass parameter in the region $-2 \leq c^3_L \leq 2$ (we determine $c^t_R$ and $c^b_R$ by the quark masses) which is more than $5 \ \sigma $ away from the measured central value  $\sigma_{ggF} = 47.8 \pm 4.0 \ \text{pb}$ \cite{ATLAS:2018doi}.  We do not consider higher absolute values of the bulk mass parameters as they would appear unnatural. 

The reason why such a large production cross section was obtained is that the top Yukawa coupling, presented as $\tilde{y}_Y- \tilde{y}_G$ in eq. \eqref{cou} of the Higgs-radion is more than $5$ times the top Yukawa coupling of the SM Higgs boson for the range $-2 \leq c^3_L \leq 2$  and its contribution adds constructively with that from the trace anomaly.  The model, within this bulk mass parameter range is phenomenologically excluded.

\section{2HDM in the bulk }
\label{2DHM in the bulk }
\noindent 

In this section we will extend the GBS scenario by adding an extra Higgs doublet in the bulk. The spacetime configuration is the same as in section \ref{GBS model}, with the metric given by equation \eqref{metric}. The action contains two scalar doublets coupled with gravity and $5D$ bulk and brane potentials. It can be written generically as
 \begin{equation}
 S =  \int d^5 x \sqrt{g} \left( - M^3R  + \frac{1}{2}g^{MN}D_M \Phi_i^\dagger D_N \Phi_i  - V(\Phi_1, \Phi_2) \right)   - \sum_{i=UV,IR} \int d^4 x  \sqrt{g_4} \ \lambda_i(\Phi_1, \Phi_2) ,
 \end{equation} 
 with $g$ is the determinant of the $5D$ metric, $M$ is the cutoff of the theory and $R$ is the Ricci scalar and where  $D_M$ is the covariant derivative containing the gauge fields.

 The radius stabilization by a Higgs doublet has been done in \cite{Geller:2013cfa,Egorov:2017vcs}, here we will closely follow their calculations while taking care of the extra Higgs in the bulk.  Electroweak symmetry breaking occurs when the Higgs doublets acquire $y$-dependent $5D$ VEVs
\begin{equation}
\Phi_1 =  \begin{pmatrix}
0 \\
\phi_1(y)
\end{pmatrix},  \quad  \Phi_2 = \begin{pmatrix}
0 \\
\phi_2(y)
\end{pmatrix}.
\end{equation}
 These VEV profiles together with the metric field $A(y)$ are determined by solving the Einstein equations
  \begin{equation}
  R_{MN} = \kappa^2 \left( T_{MN} - \frac{1}{3}g_{MN}g^{AB}T_{AB} \right),
  \end{equation}
where $\kappa^2 = 1/2M^3$ is the $5D$ Newton constant. This set of equations together with the Euler-Lagrange equations for the doublets yield the following Higgs-gravity coupled equations  \cite{ Csaki:2000zn}
 \begin{equation}
 4 A'^2- A'' = - \frac{2 \kappa^2}{3}V(\phi_1, \phi_2) - \frac{ \kappa^2}{3} \lambda_{Pl}(\phi_1, \phi_2)\delta(y) - \frac{ \kappa^2}{3}\lambda_{IR}(\phi_1, \phi_2)\delta(y-y_c),   \label{EFE1}
\end{equation}  
 \begin{equation}
 \phi_i'' = 4 A' \phi_i'  + \frac{\partial V( \phi_1, \phi_2 )}{\partial \phi_i} + \frac{\partial \lambda_{UV}(\phi_1, \phi_2)}{\partial \phi_i}\delta(y) + \frac{\partial \lambda_{IR}(\phi_1, \phi_2)}{\partial \phi_i}\delta(y-y_c)  ,     \label{EFE2}
 \end{equation}
  \begin{equation}
 A'^2 =  \frac{\kappa^2 (\phi_1'^2 + \phi_2'^2)}{12} - \frac{\kappa^2}{6} V(\phi_1, \phi_2).  \label{EFE3}
  \end{equation}
 Here primes denote derivative with respect to the extra dimension and the last equation is the zero-energy condition which, after differentiating with respect to $y$, automatically vanishes provided the other three equations are satisfied in the bulk.
 The boundary conditions are obtained by matching the delta functions 
 \begin{equation}
 A'\biggr\rvert_{y_i}  = \pm \frac{ \kappa^2}{6} \lambda_{UV,IR}(\phi_1 , \phi_2)\biggr\rvert_{y_i},  \quad  \phi_i' \biggr\rvert_{y_i} = \pm\frac{1}{2} \frac{\partial  \lambda_{UV,IR}(\phi_1 , \phi_2)}{ \partial \phi_i}\biggr\rvert_{y_i}    \label{bc1}
 \end{equation}
with the $+$ ($-$) sign for the UV (IR) boundary and orbifolding was taken into account. There is one more constraint coming from the generation of the correct gauge boson masses. Assuming the gauge boson zero modes are flat the effective VEV is given by 
\begin{equation}
v_{eff}^2 =  \int_0^{y_c} dy \left( \phi_1^2 + \phi_2^2  \right)  e^{-2 A}, \label{vev}
\end{equation}
where $v_{eff}=246 \ $GeV.

Now, we  count the parameters and constraints. From the scalar-gravity coupled system, eqns. \eqref{EFE1}-\eqref{EFE3}, we have $5$ integration constants, namely $\phi_1(0)$, $\phi'_1(0)$, $\phi_2(0)$, $\phi'_2(0)$ and $A(0)$ and there is one additional parameter, the inter brane distance $y_c$. But $A(0)$ is an irrelevant additive constant that can be absorbed by rescaling of the metric and therefore we are left with  an overall of $5$ relevant parameters and $6$ jump conditions. Thus we expect one of the boundary conditions to be fine tuned as is general in any RS-type model and this fine tuning is associated with a vanishing effective $4D$ cosmological constant. We shall discuss this counting of parameters and constraints again in the next section in the context of the superpotential generating solution.

\subsection{Superpotential}
\label{Superpotential}
\noindent

The Einstein field equations \eqref{EFE1}-\eqref{EFE3} are automatically satisfied in the bulk if we express the potential in terms of a superpotential as
\begin{equation}
V(\phi_1 , \phi_2) = \frac{1}{8} \left[ \left( \frac{\partial W(\phi_1, \phi_2)}{\partial \phi_1} \right)^2 +\left( \frac{\partial W(\phi_1, \phi_2)}{\partial \phi_2} \right)^2 \right] - \frac{\kappa^2 }{6}W(\phi_1, \phi_2)^2,  \label{SP1}
\end{equation}
provided we have 
\begin{equation}
A' = \frac{\kappa^2}{6}W(\phi_1, \phi_2),    \quad   \phi'_i = \frac{1}{2} \frac{\partial W(\phi_1, \phi_2)}{\partial \phi_i} \label{SP2}
\end{equation}
and once the appropriate boundary conditions are imposed. Another condition for \eqref{SP1} to solve the system is that the Hessian matrix of the function $W(\phi_1, \phi_2)$ is symmetric, or in other words that the second partial derivatives acting on the superpotential commute.
The advantage of using this method is that one can find a simple form for $W(\phi_1, \phi_2)$ and solve for the backreaction of the background VEVs on the metric without the assumption of small backreaction as in GBS where they used the form $A' = k + \mathcal{O}(l^2)$ in order to solve for the VEV profile.

Since the bulk Higgs doublets are fundamentals of $SU(2)_L$ we must consider a superpotential function that is bi-linear in the two doublets. For simplicity we consider the following 
\begin{equation}
W(\Phi_1, \Phi_2) =6 \frac{k}{\kappa^2} + k u_{11} \Phi_1^\dagger \Phi_1 +k u_{22} \Phi_2^\dagger \Phi_2 +  k u_{12}\left( \Phi_1^\dagger \Phi_2  + \Phi_2^\dagger \Phi_1 \right),   \label{superpotential}
\end{equation}
where the parameters $u_{ij}$ are dimensionless numbers assumed to be of $\mathcal{O}(1)$. By plugging \eqref{superpotential} into \eqref{SP1} we find that the bulk potential is given by 
 \begin{align}
V(\Phi_1 , \Phi_2) = &-6\frac{k^2}{\kappa^2}+ \bar{m}_{11}^2 \Phi_1^\dagger \Phi_1 + \bar{m}_{22}^2 \Phi_2^\dagger \Phi_2 + \bar{m}_{12}^2 \left(  \Phi_1^\dagger \Phi_2 + H.c. \right)  \nonumber \\
 &+ \lambda_1  (\Phi_1^\dagger \Phi_1)^2 + \lambda_2  (\Phi_2^\dagger \Phi_2)^2  +\lambda_3 (\Phi_1^\dagger \Phi_1)(\Phi_2^\dagger \Phi_2) + \lambda_4 (\Phi_1^\dagger \Phi_2 + \Phi_2^\dagger \Phi_1)^2    \nonumber \\
& +\lambda_6 \Phi_1^\dagger \Phi_1 ( \Phi_1^\dagger \Phi_2 + \Phi_2^\dagger \Phi_1) +  \lambda_7 \Phi_2^\dagger \Phi_2 ( \Phi_1^\dagger \Phi_2 + \Phi_2^\dagger \Phi_1),   \label{pot1}
 \end{align}
 where, with a little abuse of notation, we denote the quartic terms by $\lambda_i$, but it should be obvious that these are not the brane potentials which are defined below. 
 
The potentials on each brane are given by 
\begin{align}
 \lambda_{UV}(\Phi_1, \Phi_2) = & W(\Phi_{1}, \Phi_{2}) + V_{UV}^{2HDM}(\Phi_{1}, \Phi_{2}), \label{braneUV}
\end{align}
\begin{align}
 \lambda_{IR}(\Phi_1, \Phi_2) = &- W(\Phi_1, \Phi_2) +   V_{IR}^{2HDM}(\Phi_{1}, \Phi_{2}),  \label{braneIR}
\end{align}
where 
\begin{align}
V_i^{2HDM}(\Phi_{1}, \Phi_{2}) =& \gamma_1^{i}\left(\Phi_1^\dagger \Phi_1 - \phi_1(y_i)^2 \right)^2 +  \gamma_2^{i}\left(\Phi_2^\dagger \Phi_2 -  \phi_2(y_i)^2\right)^2  \nonumber \\
& +  \gamma_3^i \left(\Phi_1^\dagger \Phi_1+\Phi_2^\dagger \Phi_2- \phi_1(y_i)^2- \phi_2(y_i)^2 \right)^2 \nonumber \\
& +  \gamma_4^i \left(\Phi_1^\dagger \Phi_1 \Phi_2^\dagger \Phi_2 -\Phi_1^\dagger \Phi_2 \Phi_2^\dagger \Phi_1\right) +  \gamma_5^i \left( \operatorname{Re}(\Phi_1^\dagger \Phi_2)-  \phi_1(y_i)  \phi_2(y_i) \right)^2  \nonumber \\
& +  \gamma_6^i \left( \operatorname{Im} \ \Phi_1^\dagger \Phi_2 \right) ^2,
\end{align}
with $i= $UV, IR. Notice that in the boundary conditions, the brane potentials have to be evaluated at the background $\Phi_i = \phi_i$ first, and then taken derivative with respect to the profile. 
 It is remarkable that looking at solutions that have a potential of the form \eqref{SP1} immediately gives us general 2HDM in the branes and on the bulk. 

With the superpotential chosen above, the equations for the VEV profiles \eqref{SP2} become a system of homogeneous linear differential equations with constant coefficients 
\begin{equation}
\begin{pmatrix}
\phi_1' \\
\phi_2'
\end{pmatrix}  = k \begin{pmatrix}
u_{11} & u_{12} \\
u_{12} & u_{22}
\end{pmatrix}
 \begin{pmatrix}
\phi_1 \\
\phi_2
\end{pmatrix} , \label{SYS}
\end{equation}
where the solutions, satisfying the boundary conditions $\phi_i(y_c)= v_i$ can be found in Appendix \ref{2HDM} .
The warp function can be solved and written as 
\begin{equation}
A(y) = k y + \frac{\kappa^2}{12} \left[ \phi_i(y) \phi_i(y)-  \phi_i(0) \phi_i(0) \right],
\end{equation}
where the irrelevant additive constant was chosen such that $A(0)=0$ and the sum over $i$ is implicit. The backreaction of the bulk scalars on the $AdS$ background is associated with the second term in the warp function and is proportional to the square of the background vevs. 

In models with the Higgs propagating in the bulk of AdS, the localization of the Higgs profile towards the IR brane is an extra condition that needs to be fine tuned in order to generate the correct values for the electroweak gauge bosons. Therefore for the effective VEV to be  $\mathcal{O}$(TeV) we need to impose $\phi_i(0) \phi_i(0) \ll M_{Pl}^3$, see equation \eqref{vev}. Assuming $u>0$ (the case $u<0$ yields $\phi_i(0)\phi_i(0) \approx M_{Pl}^3$ and is excluded), one finds
\begin{equation}
4 u_{12}^2 =  9  +4  u_{11} u_{22} - 6 (u_{11} + u_{22})   \label{relation2}
\end{equation}
has to be tuned in order to generate a phenomenologically acceptable VEV and therefore reducing the number of free parameters. This is not surprising at all since the radius stabilization by a single $SU(2)$ doublet required the VEV profile at the Planck brane to be less than $\mathcal{O}$(TeV) \cite{Geller:2013cfa} and that condition doesn't add to the level of fine-tuning related to the cosmological constant. This is one of the fundamental differences between the Goldberger-Wise mechanism and a Higgs-doublet stabilizer.

 The hierarchy problem is resolved if we can generate the TeV scale by redshifting the Planck scale, i.e., $M_{\text{TeV}} = M_{Pl} e^{-A(y_c)}$. Solving for the warp function neglecting the term $\phi_i(0) \phi_i(0)$, we obtain, even if we assume natural values for the parameters, i.e.,  $v_i^2 \approx M_{Pl}^3$, that the contribution coming from the backreaction term above is negligible compared to that of the curvature term for the generation of the exponential hierarchy. Therefore in this model, the small backreaction approximation corresponds to
\begin{equation}
 l^2 \equiv \frac{\kappa^2(v_1^2+v_2^2)}{12} \ll 1
\end{equation} 
where $\phi_i(y_c) = v_i$.

\subsection{Scalar Perturbations}
\label{Scalar Perturbations}
In order to study the radion field we need to add the scalar perturbations about the background solution and then find the coupled Higgs-gravity Einstein equations for this fields. Following the reasoning of \cite{Csaki:2000zn} we consider the following perturbations
\begin{equation}
\Phi_i(x,y) = \begin{pmatrix}
0 \\
\phi_i(y) + \varphi_i(x,y)
\end{pmatrix},
\end{equation}
\begin{equation}
ds^2 = e^{-2A-2F(x,y)} \eta_{\mu \nu} dx^\mu dx^\nu - (1+2F(x,y))^2 dy^2.  \label{background}
\end{equation}
The linearized Einstein equations for one scalar coupled to gravity were presented in \cite{Csaki:2000zn}. For the most general case of $N$ scalars coupled minimally to gravity, see Refs.  \cite{Aybat:2010sn, George:2011tn} where the authors derived general conditions for the existence of zero-mode (massless) solutions in models with definite parity and used for particular applications of their results examples of domain- and soft-wall models.
 
 Here we concentrate on the $N=2$ case. The equations can be brought into a simpler form by writing the combination $e^{2A}\delta \mathcal{R}_{\mu \nu} + \delta \mathcal{R}_{55}$ in the bulk and integrating the $\mu 5$ equation directly.  Together with the linearized Euler-Lagrange equations, the system that has to be solved in the bulk is given by
\begin{equation}
e^{2A} \Box F + F'' - 2A' F' = \frac{2}{3} \kappa^2 \left( \phi_1' \varphi'_1 + \phi'_2 \varphi'_2 \right),  \label{linearcombination}
\end{equation}
\begin{equation}
\phi'_1 \varphi_1 + \phi'_2 \varphi_2 = \frac{3}{\kappa^2} (F' - 2 A' F),  \label{C1}
\end{equation}
\begin{equation}
e^{2A} \Box \varphi_i  - \varphi''_i +4 A' \varphi'_i + \frac{\partial^2 V}{\partial \phi_i \partial \phi_j} \varphi_j  = - 6 \phi'_i F' - 4F \frac{\partial V}{\partial \phi_i},  \label{ELEOM}
\end{equation}
where we used the background equations \eqref{EFE1}-\eqref{EFE3} to simplify. The second relation above is a constraint equation and tells us that the KK expansions of $F(x,y)$, $\varphi_1(x,y)$ and $\varphi_2(x,y)$ correspond to the same $4D$ state at each KK level so that we can write 
\begin{equation}
F(x,y)  = \sum F_n(y)h_n(x),     \quad   \varphi_i(x,y)  = \sum \varphi_{i,n}(y) h_n(x).\label{KK1}
\end{equation}

The boundary conditions are obtained by matching the delta functions in the linearized equations. Naively, the $(\mu, \nu)$ and $(5,5)$ linearized Einstein equations yield 4 BCs but as discussed in \cite{Csaki:2000zn}, they are equivalent to each other and therefore only 2 BC arise from this equations. However this $2$ boundary conditions are trivially satisfied by \eqref{C1} and the background equations, so the Einstein equations do not provide any relevant boundary conditions.
On the other hand,  the Euler-Lagrange equations give $2$ separate boundary conditions for each scalar fluctuations. These are given by 
\begin{equation}
[\varphi'_i] |_{y=y_i} =  \left[   \frac{\partial^2 \lambda_{UV, IR}}{\partial \phi_i \partial \phi_j}  \varphi_j + 2F \frac{\partial \lambda_{UV,IR}}{\partial \phi_i} \right]  \biggr\rvert_{y = y_i}   \label{BC1}
\end{equation}
where we omit the mode subindex $n$ for simplicity. As we can see, the Euler-Lagrange and Einstein equations yield a set of 4 differential equations for 3 functions of the extra dimension, namely  $F$, $\varphi_1$ and $\varphi_2$. These equations are supplemented with the boundary conditions of equations \eqref{BC1}. There are 4 integration constants and one mass eigenvalue. One integration constant corresponds to an overall normalization and the three remaining constants and the mass, together, are determined by the four boundary conditions.

The zero-mode (with $m^2=0$) solutions of the system were calculated in Ref. \cite{George:2011tn}. In that reference the authors found that in domain- and soft-wall models with $N=2$ scalars and with a bulk potential generated by a superpotential, the zero-mode solutions generally survive after imposing finite normalization and vanishing of the surface terms. Therefore they correspond to physical fields and these models are phenomenologically unacceptable. They commented that the inclusion of fundamental branes would modify the boundary conditions and could possibly render the zero-modes unphysical, making the model viable.

In this scenario, with two bulk Higgs doublets coupled to gravity, the only possible form of the bulk and brane potentials, that is consistent with $SU(2)_L$ gauge invariance, is if they are given as combinations of field bi-linears $\Phi_a^\dagger \Phi_b$, just as in \eqref{pot1}, \eqref{braneUV} and \eqref{braneIR}. More complicated expressions for the supepotential could be considered by adding terms quadratic in the fields bi-linears, i.e., $W \propto (\Phi_a^\dagger \Phi_b)^2$ but this choice of superpotential would generate non-linear ODE's for the background fields which would be more harder to solve.

For the superpotential we considered in the previous section, eq. \eqref{superpotential}, the boundary conditions for the scalar perturbations, eq. \eqref{BC1} can be written very simply as 
\begin{equation}
\varphi_i' \biggr\rvert_{y = y_i} = k u_{ij}\varphi_j \biggr\rvert_{y = y_i} + 2 k  u_{ij} \phi_j  F \biggr\rvert_{y = y_i},
\end{equation} 
and it is straightforward to show, by using \eqref{SYS}, that the zero-mode solution $F = e^{2A} A'$, $\varphi_i = e^{2A} \phi_i'$ satisfies this conditions trivially. It should be emphasized that the derivatives of the brane potentials are taken after first evaluating at the background, that is by substituting $\Phi_a  = \phi_a$ in which case the second term in \eqref{braneUV} and \eqref{braneIR} vanishes trivially. 

By inspection of the background solutions \eqref{bg1} and \eqref{bg2}, it can be seen that this functions remain finite in the interval $0<y<y_c$ and thus the normalization
\begin{equation}
N =\int dy e^{-2A} \left( \frac{2}{\kappa^2}F^2 + \varphi_i' \varphi_i' \right),
\end{equation}
would also be finite rendering the zero-mode physical and making this scenario unviable. 
Factorizing the profiles as 
\begin{equation}
F= e^{2A} \tilde{F}, \quad \varphi_i = e^{2A}\tilde{\varphi}_i,
\end{equation}
one can rewrite the bulk equations in a more compact form as 
\begin{equation}
-\begin{pmatrix}
\tilde{F}'' \\
\tilde{\varphi}_i''
\end{pmatrix}  +  \begin{pmatrix}
2A'' &  2\frac{\kappa^2}{3}(\phi_j''-A' \phi_j') \\
4(\phi_i'' - A' \phi_i')   & (4A'^2-2A'')\delta_{ij} + V_{ij}+ 2\kappa^2 \phi_i' \phi_j' 
\end{pmatrix} \begin{pmatrix}
\tilde{F} \\
\tilde{\varphi}_j
\end{pmatrix} = m^2 e^{2A} \begin{pmatrix}
                                                  \tilde{F} \\
\tilde{\varphi}_i
\                                             \end{pmatrix}, \label{D1}
\end{equation}
with the constraint equation \eqref{C1} given by 
\begin{equation}
\tilde{F}' = \frac{\kappa^2}{3} \phi_i' \tilde{\varphi}_i. \label{D2}
\end{equation}
We thus find that the argument of Ref. \cite{George:2011tn}, which applied to domain-wall and soft-wall models also applies here, leading to an unacceptable massless scalar. 

Within the present model one could consider an inert 2HDM where a discrete $Z_2$ symmetry is imposed such that only one of the scalars gets a vev profile. This case would correspond to setting the superpotential parameter $u_{12}=0$ and consequently $\tilde{m}^2_{12}=\lambda_4=\lambda_6=\lambda_7=0$, see the Appendix. Nonetheless the conclusion of this section, namely the appearance of the zero-mode, still applies and this model is ruled out. An interesting possibility would be to consider the Higgs-radion scenario and adding an extra Higgs on the TeV brane. This might avoid the appearance of a massless scalar, however generally a second doublet would increase the diphoton decay rate of the SM Higgs boson \cite{Branco:2011iw}, this would make the deviation of the original GBS model even worse. Nonetheless a full investigation of the parameter space would be interesting. Furthermore a model with 2HDM and one or more singlets would have a much more complicated potential and it is not certain that there will be a physical massless scalar in that case.

\section{Summary}
\label{Summary}
\noindent  The model of Geller, Bar-Shalom and Soni \cite{Geller:2013cfa} offers a very appealing alternative to the Goldberger-Wise stablization mechanism, in which the same field that stabilizes the radius of the extra dimension also breaks electroweak symmetry.     The model is extremely constrained, since the Higgs boson and radion are the same, and thus predictions were made for Higgs-radion production cross section and branching fractions that conflicted with the Standard Model.   Alas, during the years since their model was developed, data from the LHC has ruled out the original form of the model.    By adding bulk profiles for fermions, the parameter-space of the model is expanded, but we have shown that there are no values of these extra parameters that bring the model into agreement with LHC data.    Finally, we consider expanding the Higgs sector.     Although work of George \cite{George:2011tn} showed that soft-wall models with a potential generated by a superpotential and with more than one bulk scalar have an unacceptable physical massless scalar, we hoped that this might not apply to this model.    However, we have found that the zero mode is present and physical.    Further extending the Higgs sector would likely not change this result.    Thus, models in which the radion is a bulk Higgs doublet appear to be excluded.


\subsubsection*{Acknowledgements}


This work was funded by NSF grants PHY-1519644 and PHY-1819575.    We thank Chris Carone and Josh Erlich for helpful discussions.


\appendix

\section{Background Solution} \label{2HDM}

In this section we justify our choice of brane potentials \eqref{braneUV}, \eqref{braneIR} and give the formulas for the coefficients of the bulk potential in terms of the superpotential parameters. A derivation of the VEV profiles, equations \eqref{bg1} and \eqref{bg2}, of the system \eqref{SYS} is also given. 

For notational convenience we define the values of the VEV profiles in the branes by
\begin{equation}
\phi_i(0) \equiv \bar{\phi}_i, \quad \phi_i(y_c) \equiv v_i.
\end{equation}
Notice that given the physical input into the model, namely the bulk and brane potentials $V$ and $\lambda_{UV}$ and $\lambda_{IR}$, equation \eqref{SP1} is a non-linear partial differential equation for $W$ that has two integration constants. Equations \eqref{SP2} are first order and provide three integration constants giving a total of five integration constant as before.
The superpotential method provides the solution to the boundary value problem provided we have
\begin{equation}
\lambda_{UV}(\bar{\phi}_1, \bar{\phi}_2) =+W(\bar{\phi}_1, \bar{\phi}_2) , \quad    \lambda_{IR}(v_1, v_2)= - W(v_1, v_2),
\end{equation}
and 
\begin{equation}
\frac{\lambda_{UV}(\phi_1,\phi_2)}{\partial \phi_i} \biggr\rvert_{0} = \frac{\partial W(\phi_1,\phi_2)}{\partial \phi_i} \biggr\rvert_{0},   \quad \frac{\lambda_{IR}(\phi_1,\phi_2)}{\partial \phi_i} \biggr\rvert_{y_c} = -\frac{ \partial W(\phi_1,\phi_2)}{\partial \phi_i} \biggr\rvert_{y_c},
\end{equation}
this fact justifies our choice of brane potentials in \eqref{braneUV} and \eqref{braneIR}. It should be emphasized that brane potentials have to be evaluated at the background first, $\Phi_i = \phi_i$ and then at the orbifold fixed points.

The parameters of the 2HDM bulk potential  \eqref{pot1} are given in terms of the superpotential parameters by
\begin{equation}
\frac{\bar{m}_{11}^2}{k^2} = \frac{u_{11}^2+u_{12}^2}{8}-2u_{11}, \quad \ \frac{\bar{m}_{22}^2}{k^2} = \frac{u_{22}^2+u_{12}^2}{8}-2u_{22}, \quad \ \frac{m_{12}^2}{k^2} = u_{12}\left(\frac{u_{11}+u_{22}}{8}-2\right),
\end{equation}
\begin{equation}
\lambda_1=-\frac{k^2 \kappa^2}{6 }u_{11}^2, \quad  \lambda_2=-\frac{k^2 \kappa^2}{6 }u_{22}^2, \quad  \lambda_3 = \frac{- k^2 \kappa^2}{3}u_{11}u_{22}, \quad  \lambda_4 = -\frac{k^2 \kappa^2}{6}u_{12}^2, 
\end{equation}
\begin{equation}
\quad  \lambda_6 = -\frac{k^2 \kappa^2}{3}u_{12}u_{11}, \quad  \lambda_7 = -\frac{ k^2 \kappa^2}{3}u_{12}u_{22}
\end{equation}
therefore this choice of superpotential generates a bulk 2HDM potential with quartic interactions. Notice that quartic terms in the bulk are higher dimensional operators that one would expect to be suppressed.

 To find the solutions for the VEV profiles we notice that the system of equations can be written in matrix form as $ \textbf{ M}  \vec{ \phi}  =0$, where $\textbf{M}$ is the matrix of coefficients.
Equilibrium solutions are found by solving $\textbf{ M}  \vec{ \phi}  =0$.  We assume that $ det\bf{M} \neq  0$, so $\phi= 0$ is the only equilibrium solution. This assumption puts a contraint on the parameters of the bulk potential, namely
\begin{equation}
\det{\textbf{M}} \equiv d \equiv u_{11}u_{22}-u_{12}^2 \neq 0. \label{det}
\end{equation}
The system \eqref{SYS} can be solved easily by the inserting the ansatz
\begin{equation}
\vec{\phi}(y) = \vec{\xi} e^{r k y},
\end{equation}
with $\xi$ a constant vector. We get 
\begin{equation}
(\textbf{M} - r k \textbf{I})\vec{\xi}= 0,
\end{equation}
which is a simple eigenvalue problem. The eigenvalues are given by
\begin{equation}
r_{\pm}  =  u (2 \pm \nu ), \label{eignval}
\end{equation}
where we defined
\begin{equation}
  u \equiv \frac{u_{11}+u_{22} }{4}, \quad  \nu \equiv   \sqrt{4 - \frac{d}{u^2}}, \label{nu}
\end{equation}
and from now on we trade the parameters $\{u_{11}, u_{22},u_{12}\}$ for $\{u, \nu, u_{12}\}$.
The most general solution can be written as 
\begin{equation}
\vec{\phi}(y) = e^{2 k u y} \left( c_{+} \vec{\xi}_+ e^{k u \nu y} +  c_{-} \vec{\xi}_{-} e^{- k u \nu y} \right) , \label{sol} 
\end{equation}
where 
\begin{equation}
\vec{\xi}_{\pm}  = \frac{1}{\sqrt{2 u \nu \left[ u \nu \mp \sqrt{u^2 \nu^2-u_{12}^2}\right]}} \begin{pmatrix}
                                                                             - u_{12}  \\ \\
                                                                            \sqrt{u^2 \nu^2 -u_{12}^2} \mp u \nu
                                                                           \end{pmatrix} .  
\end{equation}
are the orthonormal set of eigenvectors and the $c_{\pm}$'s are integration constants of mass dimension $3/2$. From the discussion above, these integration constants are fixed by requiring $\phi_1(y_c) = v_1$ and $\phi_2(y_c)=v_2$ to be simultaneously satisfied. It is straightforward to find that
\begin{equation}
c_{\pm} = - e^{-(2 \pm \nu )u k y_c} \frac{u_{12}v_1 \pm v_2 (u \nu \mp \sqrt{u^2 \nu^2 - u_{12}^2})}{\sqrt{2 u \nu \left[ u \nu + \sqrt{u^2 \nu^2 - u_{12}^2}\right] }}
\end{equation}
and the background vev profiles are found to be 
\begin{equation}
\phi_1(y) = \frac{e^{2 k u (y-yc)} \left[ \left(v_1 \sqrt{\nu ^2 u^2-u_{12}^2}+u_{12} v_2 \right) \sinh (k \nu  u (y-y_c))+\nu  u v_1 \cosh (k \nu  u (y-y_c))\right]}{\nu  u},  \label{bg1}
\end{equation}
\begin{equation}
\phi_2(y) = \frac{e^{2 k u (y-y_c)} \left[ \left(u_{12} v_1-v_2\sqrt{\nu^2 u^2-u_{12}^2}\right) \sinh (k \nu  u (y-y_c))+\nu  u v_2 \cosh (k \nu  u (y-y_c))\right]}{\nu  u}.  \label{bg2}
\end{equation}


\end{document}